\documentclass[letterpaper,  
              ]{jacow}
\makeatletter%
	\ifboolexpr{bool{xetex}}
	 {\renewcommand{\Gin@extensions}{.pdf,%
	                    .png,.jpg,.bmp,.pict,.tif,.psd,.mac,.sga,.tga,.gif,%
	                    .eps,.ps,%
	                    }}{}
\makeatother

\ifboolexpr{bool{xetex} or bool{luatex}} 
 {}                                      
 {\usepackage[utf8]{inputenc}}           

\usepackage[english]{babel}			 

\usepackage[final]{pdfpages}
\usepackage{multirow}
\usepackage{ragged2e}
\usepackage{amsmath}
\ifboolexpr{bool{jacowbiblatex}}%
 {%
  \addbibresource{jacow-test.bib}
  \addbibresource{biblatex-examples.bib}
 }{}
\usepackage{tikz}
\usepackage{tikz-dsp}
\usepackage{relsize}
\usepackage{amsmath}
\usetikzlibrary{shapes,arrows,decorations.markings,chains}
\tikzstyle{vecArrow} = [thick, decoration={markings,mark=at position
	1 with {\arrow[semithick]{open triangle 60}}},
double distance=1.0pt, shorten >= 5.5pt,
preaction = {decorate},
postaction = {draw,line width=1.0pt, white,shorten >= 2.5pt}]
\tikzstyle{innerWhite} = [semithick, white,line width=1.0pt, shorten >= 2.5pt]

\DeclareMathAlphabet{\mathpzc}{OT1}{pzc}{m}{it}

\tikzstyle{block} = [draw, rectangle, 
minimum height=0.5mm, minimum width=0.5mm]
\tikzstyle{sum} = [draw, circle, node distance=1mm]
\tikzstyle{input} = [coordinate]
\tikzstyle{output} = [coordinate]
\tikzstyle{pinstyle} = [pin edge={to-,thin,black}]

\begin{document}

\title{Adaptive beam loading compensation in room temperature bunching cavities}

\author{J. P. Edelen\thanks{Now at RadiaSoft LLC, Boulder, CO, jedelen@radiasoft.org}\thanks{The authors of this work grant the arXiv.org and LLRF Workshop's International Organizing Committee a non-exclusive and irrevocable license to distribute the article, and certify that they have the right to grant this license.}, B. E. Chase, E. Cullerton, P. Varghese, Fermilab, Batavia, IL}
	
\maketitle

\begin{abstract}
In this paper we present the design, simulation, and proof of principle results of an optimization based adaptive feed-forward algorithm for beam-loading compensation in a high impedance room temperature cavity. We begin with an overview of prior developments in beam loading compensation. Then we discuss different techniques for adaptive beam loading compensation and why the use of Newton's Method is of interest for this application. This is followed by simulation and initial experimental results of this method.  
\end{abstract}

\section{Introduction}
For modern accelerators, precise control over the amplitude and phase of RF cavities is necessary. In both linear and circular accelerators, beam-loading presents a significant challenge for achieving this control, especially at higher beam intensities \cite{ref1,ref2,ref3,ref4}. Low Level RF (LLRF) feedback is capable of greatly reducing the overall impact of beam-loading
\cite{ref5,ref6}, and additional work has been done to improve feedback systems specifically targeted at beam-loading compensation
\cite{ref7, ref8, ref9, ref10}. However, the arrival of the beam produces a transient in the cavity that cannot be completely mitigated with feedback alone. As a result of the initial transient, bunches at the beginning of the beam pulse will not see the proper amplitude and phase of RF specified by the machine design. 

Some work has been done to reduce the impact of beam loading by changing the nature of the disturbance through modulation of the beam pulse train
\cite{ref11} or detuning of the RF cavity \cite{ref12}. While effective, detuning the cavity does not completely mitigate the beam-loading transient, and changing the nature of the pulse train is not optimal for a linear accelerator with tight requirements on the beam parameters. Feed-forward schemes \cite{ref13,
ref14} are attractive because in principle they can completely mitigate the beam loading disturbance from the RF cavity. There has been some development in this area, including the use of beam diagnostics to estimate the beam loading seen by the cavities and then use this information to apply the appropriate feed-forward correction during the RF pulse \cite{ref15,
ref16}. The use of external diagnostics however, is not as reliable as an adaptive feed-forward algorithm within the LLRF system. These methods seek to minimize the beam-loading disturbance through iterative adjustment of the feed-forward correction based on the measured error between the cavity field and set-point \cite{ref17,
ref18, ref19, ref20}. The adaptive algorithms tend to be computationally simple and can easily be implemented on either the LLRF microprocessor or on a FPGA. One drawback of this method is that the iteration process generally requires a finite number of pulses to converge, resulting in some beam pulses being out of tolerance while the algorithm is converging. In an effort to minimize the number of pulses needed to reduce the beam loading disturbance, we investigate the use of Newton's method in an adaptive feed-forward algorithm. 

In this paper we will show the design, simulation, and proof-of-principle experimental results for this algorithm applied to a high-impedance bunching cavity currently in place in the the Proton Improvement Plan II injector test (PI-Test) linear accelerator. We will begin with an overview of the bunching cavity and characterization of the effects of beam loading in the cavity. We then introduce an adaptive controller that utilizes Newton's method in order to minimize the beam-loading disturbance and show our experimental results.

\section{Overview of the bunching cavity}
The proposed upgrade to the Fermilab accelerator complex referred to as PIP-II is currently under development in order to meet the requirements for the next generation of neutrino experiments
\cite{ref21}. In order to assess the feasibility of this objective and address the technical gaps in current accelerator technology, the PIP-II injector test
\cite{ref22} is being commissioned. The warm front-end of the injector test contains four normal-conducting RF cavities, an RFQ that focuses the beam and accelerates it to 2.1 MeV, and three bunching cavities to provide additional longitudinal focusing. The bunching cavities are room-temperature quarter-wave resonators with a design frequency of 162.5 MHz. Currently, one bunching cavity is in place in the beam-line. This cavity has a $Q_\mathrm{L}$ of 5280 and an operating voltage of 85 kV. Table 1 shows the figures of merit for the bunching cavity, as measured during its initial check-out and conditioning
\cite{ref23}. 

\begin{table}[h]
\caption{Figures of merit for the PI-Test bunching cavity}
\centering 
\begin{tabular} {c|c}
Figure of merit & Value \\ \hline
$f_\mathrm{0}$ & 162.5 [MHz] \\
$Q_\mathrm{L}$ & 5280 \\ 
$\beta$ & 0.9 \\ 
$R_s$ & 7M$\Omega$ \\
$T$ & 0.675  \\
$\Delta f_0 /\Delta T_\mathrm{H_2O}$ & -1.65 [kHz/$^o$F]\\
$\Delta f_0/\Delta P_\mathrm{rf} $& -9.5 [kHz/kW] \\
$\Delta T_\mathrm{H_2O} $ & $\pm 2^o$F\\
$V_\mathrm{cavity} $ & 85 kV  \\ 
$E_\mathrm{peak} $ & 1.65 MV/m \\ 
\end{tabular} 
\end{table} 

The ion source for PI-Test is designed to produce up to 10 mA of beam current, with a nominal output of 5 mA. This results in a beam-induced voltage in the cavity of 24 kV and 12 kV respectively. Because of the relatively high beam voltage relative to the cavity voltage, the cavities are subject to heavy beam loading. Under normal operating conditions, the beam phase relative to the RF is -90$^\circ$, therefore the beam loading on the cavity is seen primarily as a phase disturbance. For a 5-mA beam the phase disturbance is 7.9$^\circ$, as shown in Figure 1. 

\begin{figure}[h]
\centering
\includegraphics[width=0.5\textwidth]{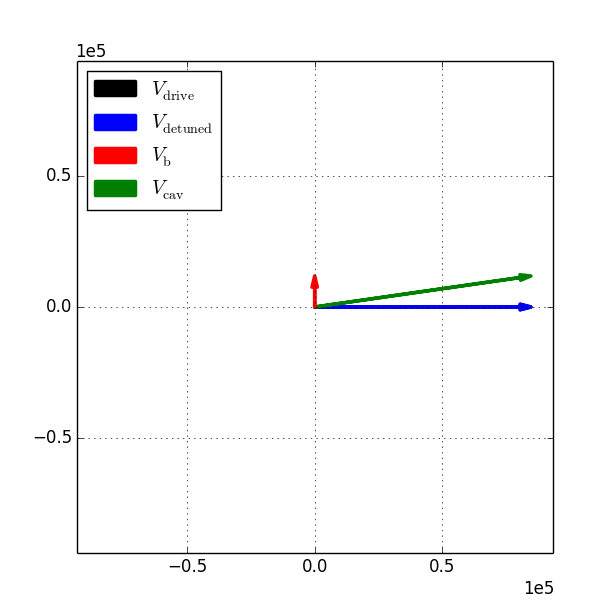}
\caption{Phasor diagram of beam-loading disturbance for nominal current with no detuning. Note that because the beam is H-, we have a positive kick in the phase.} 
\end{figure} 

The feedback system will compensate for a large portion of this disturbance; however during the first $20\mu s$ of beam there will be an uncompensated transient in the cavity. This transient impedes the cavities ability to properly focus the beam and introduces unwanted energy gain during the first 20 $\mu$s of beam.

\section{Adaptive feed-forward algorithm}
For a linear system Newton's method is guaranteed convergence in two iterations 
\cite{ref26}. Because beam loading for an ideal beam is linear, Newton's method represents the fastest possible means for optimizing the beam-loading compensation. The choice of objective function is important for any optimization problem, and for the bunching cavity we choose to minimize the integrated in-phase and quadrature disturbances seen by the LLRF system, Equation 1.

\begin{equation}
f(A_n)=\int_{t_0}^{t_{stop}}(V_{cav}(t)-V_{set}(t))dt
\end{equation} 

Through the application of Newton's method we can determine the needed feed forward correction $A_n$ as a function of the integrated disturbance, $f(A_n)$, and the step size $\Delta A_n$, Equation 2. 

\begin{equation} 
A_{n+1}=A_n- {f(A_n) \Delta A_n \over f(A_n+\Delta A_n)-f(A_n)}
\end{equation} 

Here $A_n$ is the feed-forward compensation at the $n^{th}$ iteration, $\Delta A_n$ is the iteration step used to compute the local slope, and $f(A_n)$ is the integrated error for either the in-phase or quadrature signal along the RF pulse. In order to determine the stability of the algorithm, we invoke the bounded-input bounded-output stability criteria
\cite{ref27}. For Equation 2 to have an unbounded output either the numerator must have some exponential growth term or the denominator must go to zero. If the integrated disturbance (Equation 5) is bounded then the numerator in Equation 6 must also be bounded. Therefore the stability of Equation 1 is satisfied by the requirement that the LLRF feedback system be stable and that the step size be greater than zero. Because these two criteria are necessarily satisfied, we can assert that this algorithm is stable. 

Using a base-band model of the LLRF system, we can simulate the performance of the algorithm using changes in the beam phase as the disturbance that triggers the optimization routine. This is representative of the phase scans performed during beam characterization and machine tune-up.  Figure 2 shows the results of the simulated phase scan. 

\begin{figure}[h] 
   \centering
   \includegraphics[width=0.5\textwidth]{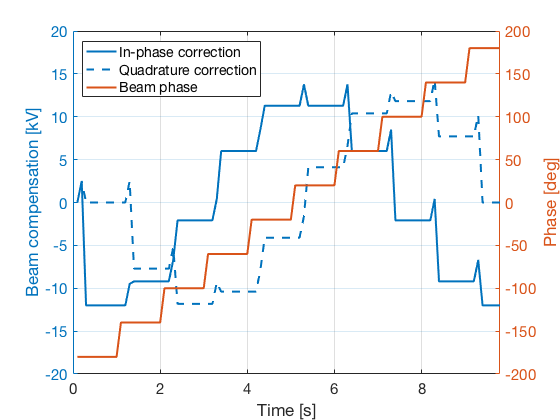} 
   \caption{Blue is the output of the optimization routine and red is the beam phase. The algorithm is initialized with no compensation.}
   \label{fig:figure_6}
\end{figure}

Here we see that after the phase of the beam changes, the algorithm is triggered to re-optimize the beam-loading compensation. For each change in phase the beam-loading compensation converges in the aforementioned minimum number of steps. For the purposes of this simulation the time scale assumes that the RF pulses occur at a repetition rate of 10Hz.

\section{Implementation on the LLRF front-end}
The algorithm was implemented on the LLRF front-end controller in order to test its performance on the machine. The controller is a VME 5500 board running VXWorks 6.4 with a data acquisition cycle rate of 10 Hz. Because this particular controller is interfacing with two FPGA cards that each control a different cavity, writing an update to the FPGA every DAQ cycle would cause the system to freeze up. To mitigate this, the two controllers are updated in an alternating fashion. This gives the state flow described in Figure 3.

\begin{figure}[h!] 
   \centering
   \includegraphics[width=0.5\textwidth]{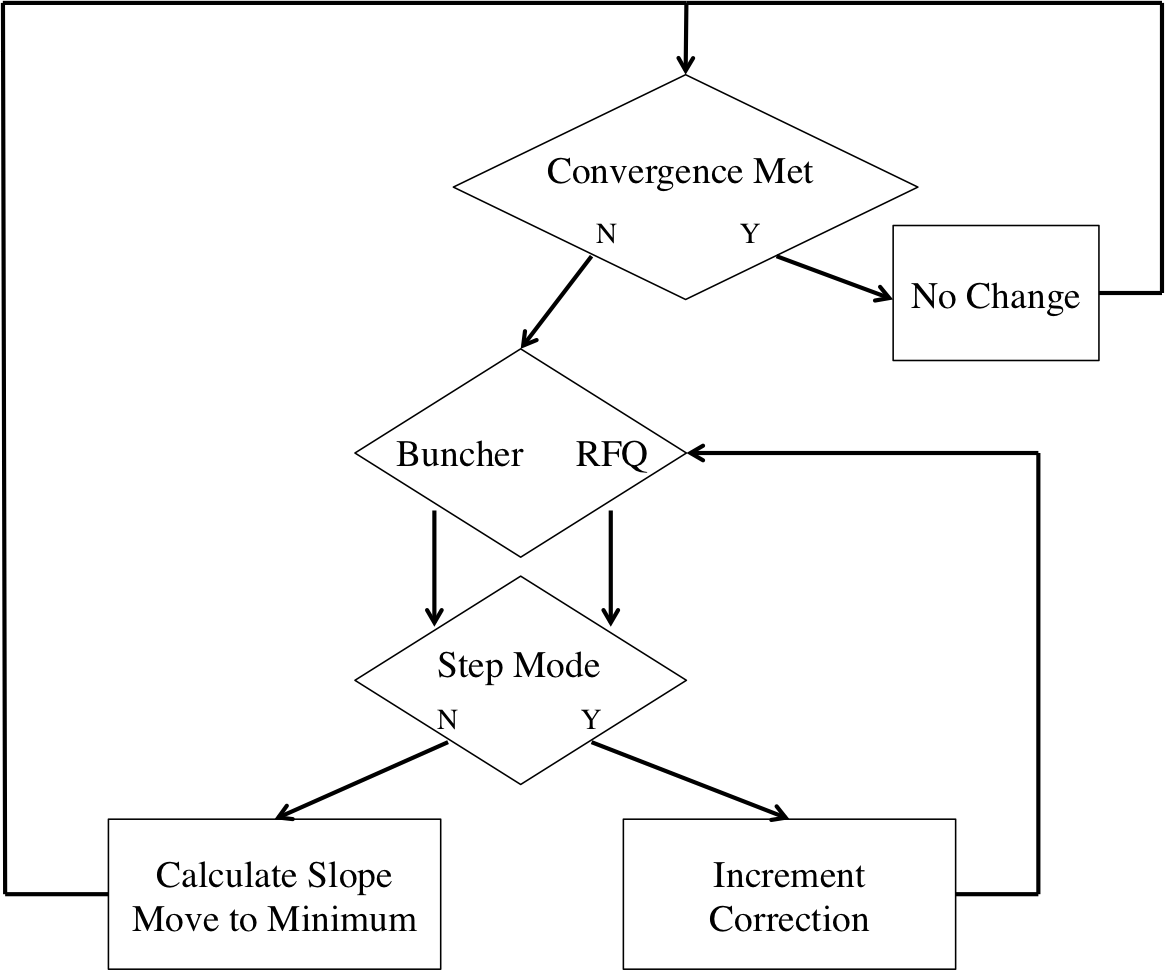} 
   \caption{Block diagram of the implemented algorithm.}
   \label{fig:figure_7}
\end{figure}

The error function defined in Equation 1 is computed numerically from the in-phase and quadrature signals read back to the front end, (Equation 3). Here we assume the cavity has reached a steady state before the arrival of the beam. Normally the error would be computed relative to the in-phase and quadrature set-points stored in the LLRF controller, however, additional terms in the controller and noise add offsets to the measurements. Therefore, in order to ensure proper offset subtraction, we use the value in the data table just before our beam-compensation window as the set-point. 

\begin{equation}
\begin{split}
I_\mathrm{tot}= \sum_{n=n_0}^{n_0+n_\mathrm{beam}} (I[n] - I[n_0-2])\\
Q_\mathrm{tot} = \sum_{n=n_0}^{n_0+n_\mathrm{beam}}(Q[n] - Q[n_0-2])
\end{split}
\end{equation}
  
Here $n_0$ is the start of the beam-loading compensation pulse, and $n_\mathrm{beam}$ is the length of the beam-compensation pulse. Additionally, due to the large dynamic range and the relatively small error signals present in the beam-loading error measurement we chose not to multiply Equation 3 by the time step. Because the time step is not changing pulse-to-pulse it should not change the fundamental behavior of the algorithm. 

\section{Beam loading measurements} 
Prior to testing the algorithm, we performed several measurements of the beam loading disturbance using the signals computed in the previous section. These measurements are important for characterizing the beam-loading and also for validating our simulations. The first measurement scanned the phase of an ideal beam-like disturbance driven by the LLRF system. Figure 4 shows this disturbance as a function of its drive phase. 

\begin{figure}[h] 
   \centering
   \includegraphics[width=0.5\textwidth]{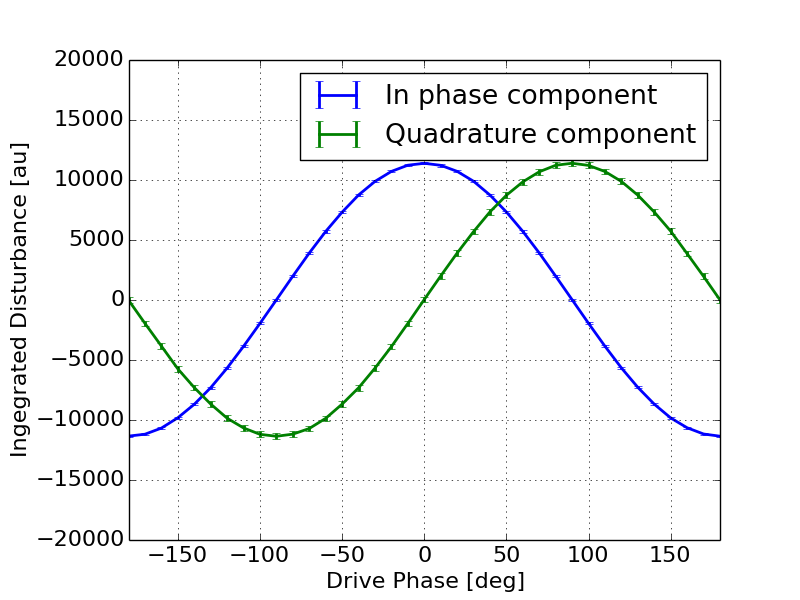} 
   \caption{Measured beam disturbance from ideal pulse driven from the LLRF system}
   \label{fig:figure_8}
\end{figure}

Figure 4 shows what one would expect from this ideal disturbance. The in-phase and quadrature components of the error signal have the same amplitude and are precisely 90$^\circ$ out of phase. Next we measured the beam-loading disturbance from a 5-mA, 50 $\mathrm{\mu}$s beam as a function of the beam phase, Figure 5.

\begin{figure}[h] 
   \centering
   \includegraphics[width=0.5\textwidth]{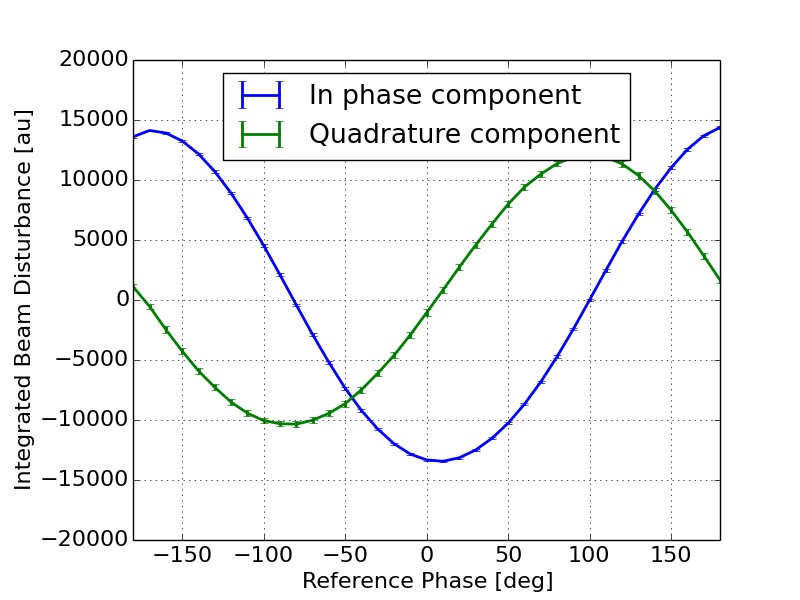} 
   \caption{Measured beam disturbance}
   \label{fig:figure_9}
\end{figure}

Here we see that the disturbance measured by the beam does not match that of the ideal disturbance driven by the LLRF system. Not only are the in-phase and quadrature components not the same magnitude, but analysis of these two waveforms show that they are approximately 2.1$^\circ$ from being orthogonal. Put another way, there is 92.1$^\circ$ difference between the two waveforms.   

\section{Algorithm performance} 
To test the algorithm we scanned the phase of the beam relative to the RF by adjusting the reference phase. During these measurements we recorded the in-phase and quadrature corrections as well as the integrated errors. The reference phase was varied from -180$^\circ$ to 180$^\circ$ in 10$^\circ$ increments. Each phase step was held for a fixed length of time to allow for the algorithm to converge. This time is referred to as our convergence window. The phase scan was performed for two different convergence windows: 15 seconds and 30 seconds. Prior to the phase scan we tested the sensitivity of the algorithm to noise and settled on 1000 counts as the maximum error allowable for convergence. The step size of the algorithm was 150 counts. The final beam-loading correction for both the 15-second window and the 30-second window are shown in Figure 6.  During this scan the the cavity was placed at its nominal settings of -90$^\circ$ and 85 kV.  

\begin{figure}[h] 
   \centering
   \includegraphics[width=0.5\textwidth]{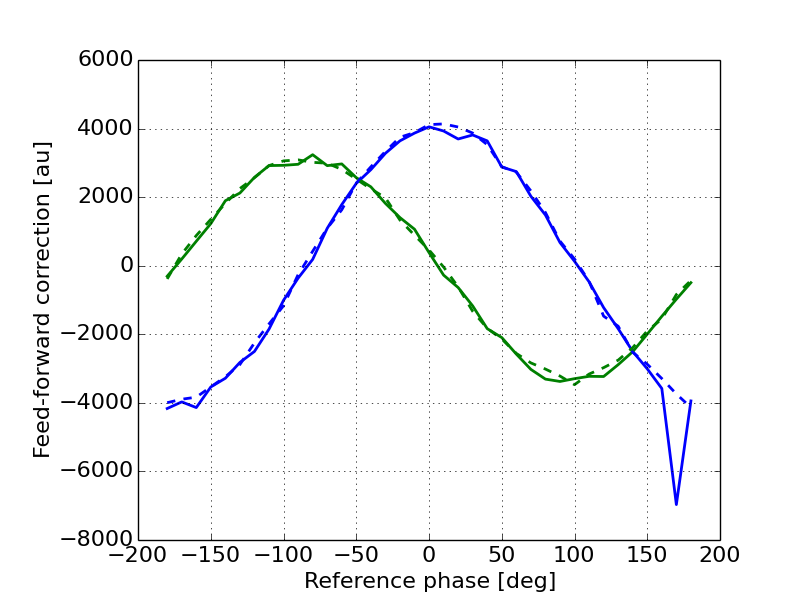} 
   \caption{Correction as a function of beam phase. The dashed line is the 30-second window, and the solid line is the 15-second widow. Note that for the 30-second window, the final convergence was a bit better than the 15-second window, resulting in slightly better performance.}
   \label{fig:figure_10}
\end{figure}

Here we see that the shape of the correction in both the in-phase and quadrature components matches the shape of the disturbance measured in Figure 5. This indicates  that the algorithm is insensitive the non-ideal nature of the beam loading for this particular system. While performing the phase scan we observed that some phases took the minimum number of steps to converge while others took many steps to converge. Figure 7 shows an example of a change in beam phase that took the minimum number of steps, while Figure 8 shows an example of a change in beam phase that resulted in many oscillations before finally converging.

\begin{figure}[h] 
   \centering
   \includegraphics[width=0.5\textwidth]{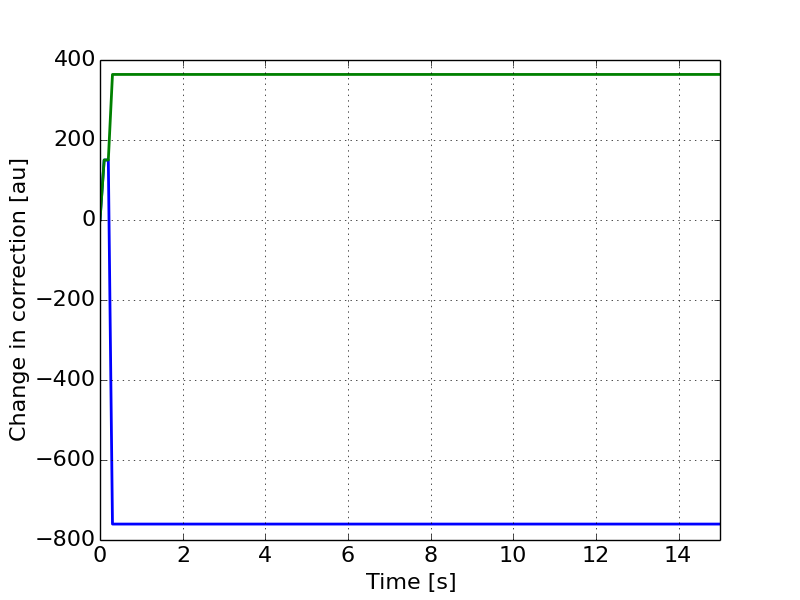} 
   \caption{Demonstration of ideal convergence time with the adaptive algorithm: in-phase (blue), quadrature (green) }
   \label{fig:figure_11}
\end{figure}

\begin{figure}[h] 
   \centering
   \includegraphics[width=0.5\textwidth]{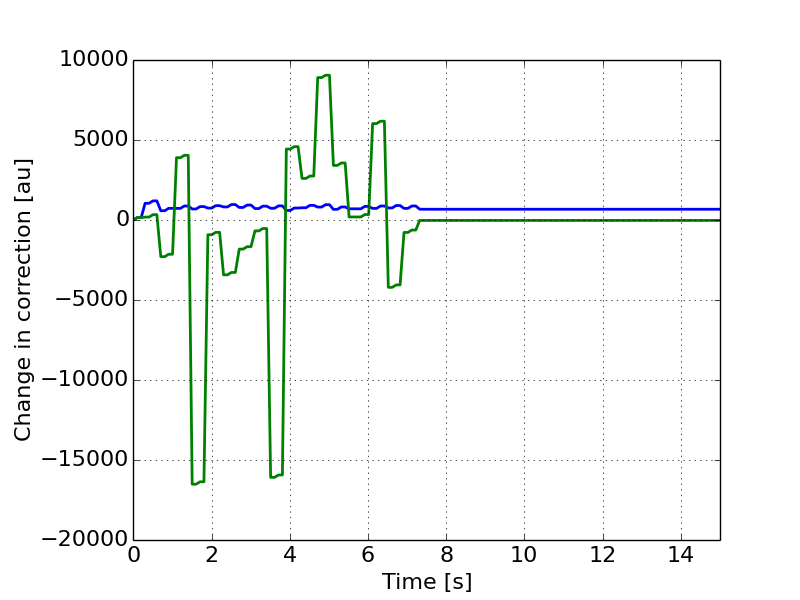} 
   \caption{Example of non-ideal convergence time with the adaptive algorithm: in-phase (blue), quadrature (green)}
   \label{fig:figure_12}
\end{figure}

After investigation of the signals in the LLRF system we discovered a small amplitude modulation being generated in the amplifier. This amplitude modulation can create errors in the local slope calculation leading to oscillations in the feed-forward correction and therefore decreased performance of the algorithm.

\section{Conclusions} 
We have tested the use of a Newton's Method based algorithm for beam-loading compensation in a high-impedance cavity. While in some cases the algorithm converges in the minimum number of iterations there were some iterations that took considerably longer to converge due to amplitude modulation from one of the amplifiers. Here a simple integrator would preform much better. In systems where convergence time is critical, this method could be applied but care should be taken in the implementation to avoid unstable behavior due to noise or other disturbances in the RF system.

\end{document}